\documentclass[preprint,12pt]{elsarticle}

\usepackage[T1]{fontenc}
\usepackage{amssymb}
\usepackage{float} 
\usepackage{caption}
\usepackage{adjustbox}
\usepackage{graphicx}%
\usepackage{multirow}%
\usepackage{amsmath,amssymb,amsfonts}%
\usepackage{booktabs} 
\usepackage{rotating}

\usepackage[breaklinks]{hyperref}

\begin{document}

\begin{frontmatter}


\author{Suchithra Kunhoth\corref{cor1}\fnref{label1}}
\ead{suchithra@qu.edu.qa}
\cortext[cor1]{Corresponding author}

\title{Computational Methods for Breast Cancer
Molecular Profiling through Routine
Histopathology: A Review}


\affiliation[label1]{organization={Computer Science and Engineering, Qatar University},
            city={Doha},
            country={Qatar}}

\author[label1]{Somaya Al- Maadeed}
\author[label1]{Younes Akbari}
\author[label2]{Rafif Al Saady}

\affiliation[label2]{organization={College of Medicine, QU Health, Qatar University},
            city={Doha},
            country={Qatar}}

\begin{abstract}
Precision medicine has become a central focus in breast cancer management, advancing beyond conventional methods to deliver more precise and individualized therapies. Traditionally, histopathology images have been used primarily for diagnostic purposes; however, they are now recognized for their potential in molecular profiling, which provides deeper insights into cancer prognosis and treatment response. Recent advancements in artificial intelligence (AI) have enabled digital pathology to analyze histopathologic images for both targeted molecular and broader omic biomarkers, marking a pivotal step in personalized cancer care. These technologies offer the capability to extract various biomarkers such as genomic, transcriptomic, proteomic, and metabolomic markers directly from the routine hematoxylin and eosin (H\&E) stained images, which can support treatment decisions without the need for costly molecular assays. In this work, we provide a comprehensive review of AI-driven techniques for biomarker detection, with a focus on diverse omic biomarkers that allow novel biomarker discovery. Additionally, we analyze the major challenges faced in this field for robust algorithm development. These challenges highlight areas where further research is essential to bridge the gap between AI research and clinical application.
\end{abstract}

\begin{highlights}
\item Explores research works intended to extract omic as
well as non-omic biomarkers for breast cancer from routine histopathological images
\item Identify the existing computer vision and AI based methods used in this context
\item Highlight the challenges that hinder the research progress and impact the clinical deployment
\end{highlights}


\end{frontmatter}


\section{Introduction}\label{sec1}

Breast cancer is the most frequently diagnosed cancer among women globally, accounting for 7\% of all cancer-related deaths worldwide \cite{1}. According to 2020 statistics, breast cancer was the leading cause of cancer-related deaths among the Qatari population, contributing to 18.47\% of such fatalities \cite{91}. While early detection through screening plays a crucial role in reducing mortality and morbidity, the development of effective treatment strategies is equally important. A significant challenge in managing breast cancer arises from its heterogeneity, which is driven by diverse genetic alterations in individual patient. This variability contributes to differences in patient prognosis, treatment response, and outcomes \cite{2}. However, in recent years, precision medicine has largely been utilized to gain more insights into the molecular level understanding of breast cancer. Through advanced molecular profiling techniques, precision medicine seeks to achieve the goal of personalized cancer treatment \cite{3}.

Molecular profiling techniques attempt to identify the underlying characteristics of cancer cells by examining specific molecules such as Deoxyribonucleic acid (DNA), Ribonucleic acid (RNA), Proteins, Metabolites, etc., \cite{4}. Multiple omics technologies serve to generate vast amount of data about specific type of molecules. This includes molecular characterization using genomics (DNA), transcriptomics (RNA), proteomics (protein), metabolomics (metabolites), etc. Data generated from the omic technologies (eg:- Next Generation Sequencing for Genomics and Transcriptomics) are analysed to identify specific biomarkers that are related to the disease occurrence, prognosis as well as treatment decisions. Further to the validation, these biomarkers are approved to be used in the clinical settings \cite{5}. The identification of novel biomarkers has proven effective in distinguishing cancer patients, enabling the implementation of personalized treatment plans and improving patient care. Techniques like immunohistochemistry (IHC) and fluorescence in situ hybridization (FISH), which are already established in clinical laboratories, serve as fundamental tools in molecular profiling for detecting individual biomarkers. These methods are useful for identifying the presence of specific proteins or immune-related biomarkers. However, to analyze an entire set of proteins, proteomic techniques, such as mass spectrometry, are necessary. Although new biomarkers are continuously being discovered \cite{6}, Table \ref{tab1} provides a summary of the most commonly used omic and non-omic biomarkers in breast cancer.

\begin{table}[!h]
\small
\caption{Commonly used biomarkers in breast cancer} \label{tab1}%
\begin{tabular*}{\textwidth}{@{\extracolsep\fill}p{2cm}p{4cm}p{7cm}}
\toprule
Biomarker  & Details  & Significance\\
\midrule
ER \cite{86}    & Hormone receptor status for Estrogen Receptor   & Determine cancer's likeliness to respond to hormone therapy (e.g., tamoxifen or aromatase inhibitors) \\
PR \cite{86}  & Hormone receptor status for Progesterone Receptor   & Determine cancer's likeliness to respond to hormone therapy (e.g., tamoxifen or aromatase inhibitors) \\
HER2 \cite{86}  & Human Epidermal growth factor Receptor 2  & Guides the use of HER2-targeted therapies such as trastuzumab \\
Ki-67 \cite{86}  & Proliferation Marker ki-67  & Associated with cell proliferation, Seen in Aggressive tumors, High Risk of recurrence \\
BRCA1/ BRCA2 \cite{86}  & Genetic Biomarkers that may undergo mutations & Associated with hereditary breast cancer, guide the use of PARP inhibitors \\
PIK3CA \cite{86}  & May undergo genomic alterations  & Targeted by specific inhibitors \\
TP53 \cite{86}  & Tumor Suppressor Gene  & Aggressive Tumor behaviour \\
PDL1 \cite{86}  & Immunotherapy Marker  & Expression is used to identify patients with TNBC who may benefit from immune checkpoint inhibitors \\
\multirow{3}{=}{Gene Expression Signatures \cite{86}} & Oncotype DX: 21-gene signature test & Predict the likelihood of breast cancer recurrence \\ 

    & MammaPrint: 70-gene signature test & Classifies breast cancer patients into low-risk or high-risk categories for recurrence \\ 

    & PAM50: 50-gene signature test & Assesses the risk of distant recurrence and helps classify breast cancer into intrinsic subtypes \\ 
\bottomrule
\end{tabular*}
\end{table}

The detection of non-omic biomarkers, such as ER, PD-L1, and Ki-67, often necessitates specific staining procedures in immunohistochemistry (IHC) to make these biomarkers visible to the pathologist. In addition, the type of stain is specific for each biomarker. The staining intensity and quality can vary depending on factors like antibody type, staining protocols, and even lab-to-lab variability. This may lead to inconsistency among the result interpretations due to the intra and inter observer variability between pathologists. Concerning the omic biomarkers, the methods involved are highly expensive and time consuming. Since it is not routinely performed in all labs, a large group of people can miss the benefits of molecular profiling \cite{7}. The rapid progress in artificial intelligence and deep learning paved the way to provide automated techniques for the medical image analysis. This enabled to provide efficient algorithms to accomplish many tasks such as cancer detection \cite{84}, classification, risk prediction \cite{8} etc. Recent studies have demonstrated that it is feasible to predict one or more molecular biomarkers directly from histopathology images \cite{87}. 

The paper is organized as follows: Section 2 outlines the motivation behind the research and highlights the major contributions of our work. Section 3 introduces some relevant review articles and reveals how our article is different from those works. Section 4  and 5 explores the review of computational methods using H\&E images that could potentially replace the traditional non omic and omic biomarker extraction respectively. Section 6 analyzes the significant aspects of the discussed computational methods, along with identifying the future prospects. Finally, section 7 brings the review article to a conclusion.

\section{Motivation and contribution}\label{sec2}
\begin{figure}[h]
\centering
\includegraphics[width=0.9\textwidth]{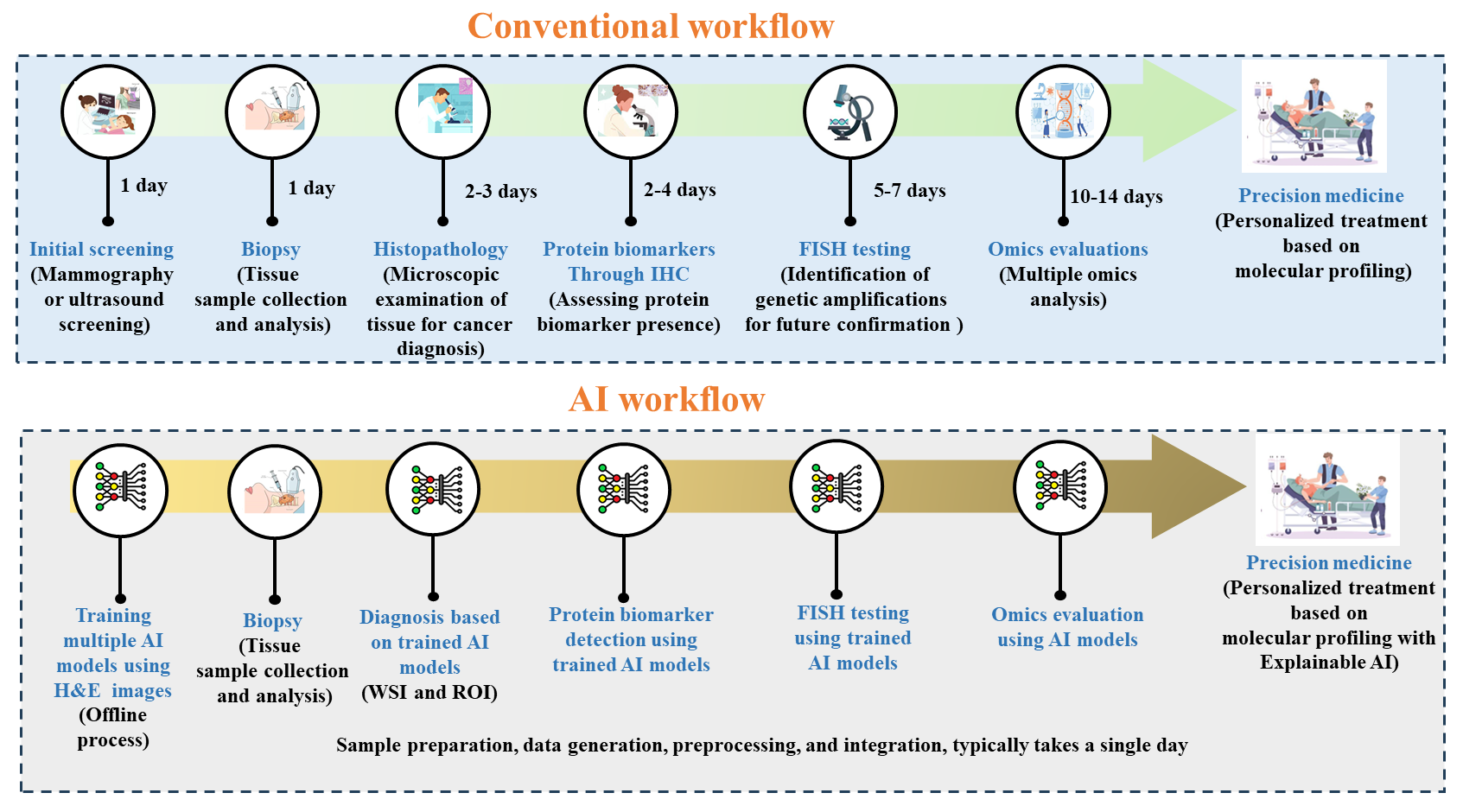}
\caption{Conventional workflow in breast cancer management from screening to personalized treatment}\label{fig1}
\end{figure}

The standard clinical pathway that begins with the detection and classification of the tumor and progresses to the use of molecular techniques that inform therapeutic decision-making is illustrated in Figure \ref{fig1}. While individual omics like genomics is used to some extent in clinical practice, comprehensive multi-omic analysis is primarily in the exploratory phase. However, as technology advances and costs decrease, it holds promise for becoming a routine part of breast cancer diagnosis and treatment planning, enhancing the precision of treatment decisions. The routine histopathology slides are usually stained with hematoxylin and eosin, which is referred to as the H\&E slides. These slides need to undergo the specific staining process for the concerned biomarker detection through IHC. Many of the initial researchers have worked on the prediction of biomarkers from the whole slides images captured from these IHC slides. This staining inconsistency can affect how well an algorithm detects positive Ki-67 or ER cells. High variability in staining protocols makes algorithms less generalizable unless explicitly trained on diverse datasets covering multiple staining techniques. H\&E is a more standardized staining method, widely used across institutions with fewer significant variations compared to Ki-67. This consistency allows H\&E-based algorithms to perform more robustly across diverse samples and settings. Subsequent research began predicting biomarkers from digitized H\&E images. Although identifying and classifying histologic and morphologic features from H\&E slides was once limited to what pathologists could observe, this technique has unlocked access to information not visible to the naked eye. Additionally, it has become possible to predict multiple omic biomarkers directly from whole-slide H\&E images. Traditionally, identifying each biomarker often required additional tissue slices. This may result in loss of some tumor markers especially with small biopsies. The automated analysis of digitized H\&E slides for biomarker identification helps alleviate many of these burdens on the medical system. This highlights the need to review the literature on biomarker identification from routine histopathological images of breast cancer biopsy samples.

The contributions of the work are:

\begin{itemize}
\item An in depth examination of research works aiming to extract omic as well as non-omic biomarkers from routine histopathological samples through digital image analysis.

\item To conduct a review properly categorized by the commonly used breast cancer biomarkers and the relevant computer vision techniques employed to predict them. 

\item The work aims to figure out the recent contributions of artificial intelligence towards achieving the goals of precision medicine for breast cancer.
\end{itemize}

\section{Existing Studies and our work}\label{sec3}

\begin{figure}[!h]
\centering
\includegraphics[width=0.9\textwidth]{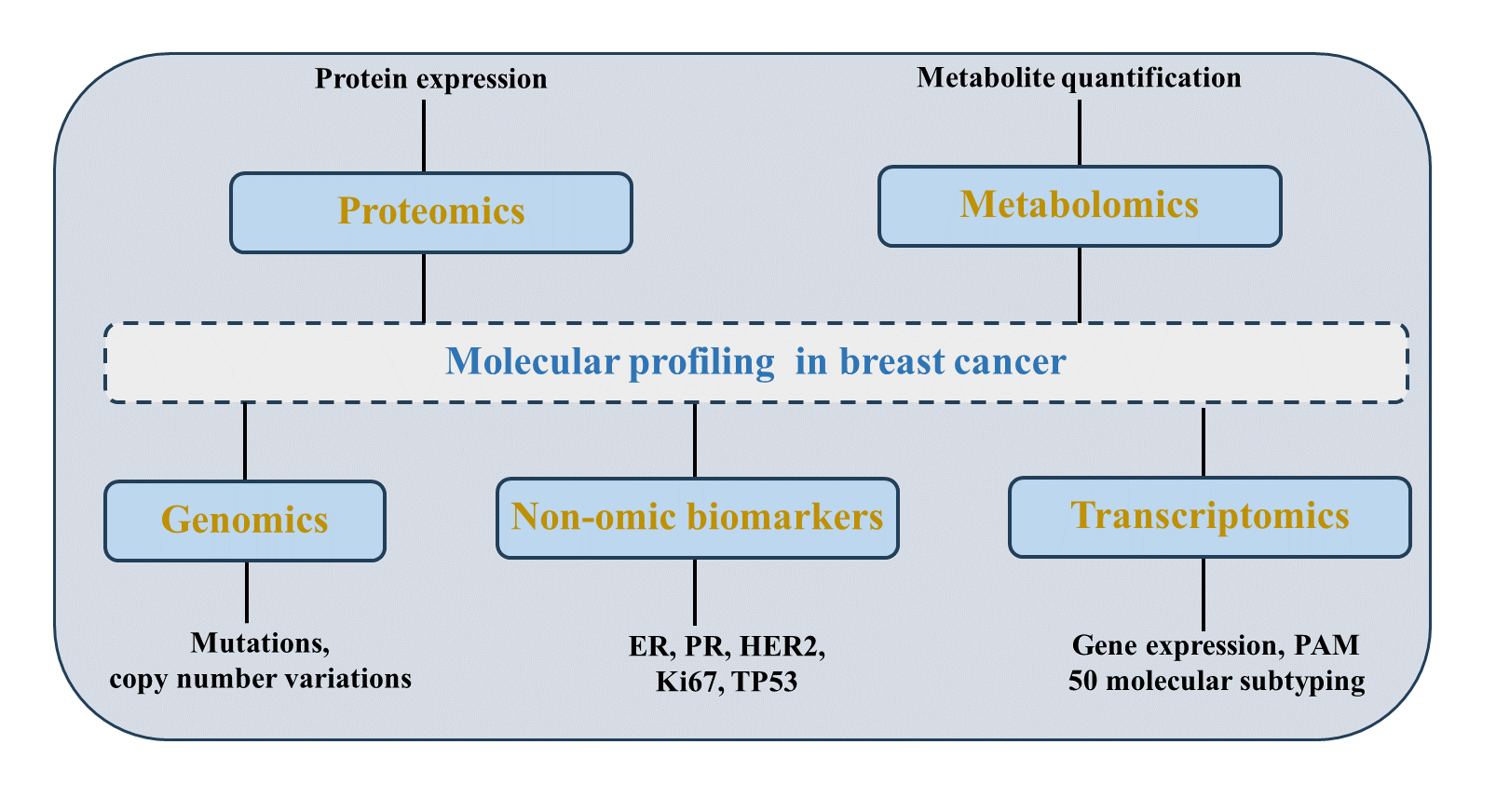}
\caption{Overview of molecular profiling approaches in breast cancer, each contributing to personalized treatment insights}\label{fig2}
\end{figure}

In recent years, several review articles have addressed the applications of artificial intelligence in the diagnosis and management of breast cancer. Artem et al. \cite{9} conducted a study that explores the comprehensive applications of artificial intelligence in histopathology for various cancers. This includes the use of AI for predicting disease outcomes and genetic alterations from digitized histopathology slides. Another recent review \cite{10} specifically focuses on breast cancer and deep learning. Contrary to what the title suggests, this work does not limit itself to articles based solely on histopathology images; it also encompasses applications related to radiology and omic data. The review in \cite{11} concentrates on computational methods for identifying biomarkers from multi-omics data and presents several available omic data repositories. Another review \cite{12} discusses the detection of different cancer biomarkers through digital image analysis. Gauhar et al. provided a comprehensive overview of HER2 automation classification algorithms published over the last decade \cite{13}. The review article \cite{14} discusses deep learning applications in cancer pathology, covering not only fundamental applications but also advanced image analysis tasks such as molecular feature assessment and survival prediction from digital images. The prediction of molecular biomarkers from H\&E images was reviewed in \cite{15}. This work addresses biomarkers for approximately ten different cancer types and includes the computational methods for the extraction of protein biomarkers, genomic subtypes, individual gene expression, and molecular alterations more generally. However, it lacks clarification on the medical aspects relevant to each of those approaches. Specifically, the different omic data types, such as genomic, transcriptomic, and proteomic data, and their clinical relevance were not fully addressed. Additionally, the specific datasets used for these analyses, the type of data they contain, performance and how they align with medical aspects were not described in sufficient detail. This omission leaves a gap in understanding how computational methods are connected to real-world clinical applications and outcomes. The comparison of our proposed research with the existing reviews is shown in Table \ref{tab2}. Our review focuses solely on methods for identifying breast cancer biomarkers from digital H\&E slide images, with a particular emphasis on their medical relevance. The different aspects of molecular profiling covered in this study is illustrated in Figure \ref{fig2}. All the individual molecular biomarkers are classified as non-omic biomarkers, while other omics categories are addressed separately.

\begin{table}[h]
\small
\caption{Comparison of proposed research with recently published review articles}\label{tab2}%
\begin{tabular*}{\textwidth}{@{\extracolsep\fill}p{2cm}p{1.5cm}p{1.5cm}p{1.5cm}p{1.5cm}p{1.5cm}p{2cm}}
\toprule
Ref. & Biomarkers from Images   & H\&E images only & Breast Cancer Focus & Dataset Details &  Computational-Clinical Link\\
\midrule
\cite{9} &  $\checkmark$   & $\times$  & $\times$  & $\times$ & $\checkmark$ \\
\cite{10} & $\checkmark$   & $\times$ & $\checkmark$ & $\times$ & $\checkmark$ \\
\cite{11} & $\times$ & $\times$ & $\times$ & $\checkmark$ & $\checkmark$ \\
\cite{12} & $\checkmark$ & $\times$ & $\times$ & $\times$ & $\checkmark$  \\
\cite{13} & $\checkmark$   & $\times$ & $\checkmark$ & $\checkmark$ & $\checkmark$  \\
\cite{14} & $\checkmark$  & $\checkmark$ & $\times$ & $\checkmark$ & $\checkmark$  \\
\cite{15} & $\checkmark$  & $\checkmark$  & $\times$ & $\times$ & $\times$  \\
Our review & $\checkmark$  & $\checkmark$  & $\checkmark$  & $\checkmark$  & $\checkmark$  \\
\bottomrule
\end{tabular*}
\end{table}

\section{Computational Techniques Replacing Traditional IHC/FISH-Based Biomarker Extraction}\label{sec4}

IHC is the gold standard method for the detection of protein biomarkers. Protein biomarkers play a significant role in breast cancer diagnosis, prognosis, treatment choice and treatment response \cite{16}. The established biomarkers such as ER, PR, HER2 are commonly recommended for use in clinical laboratories. If the result of the IHC turned out to be equivocal for HER2, further testing through fluorescent in situ hybridization analysis (FISH) is needed to confirm the gene amplification \cite{88}. Ki67 is another protein marker associated with cell proliferation \cite{89}. IHC measures the expression level of this biomarker as the percentage of positively stained cells. Ki67 serves as an important tool to differentiate between the luminal A and luminal B subtypes of breast cancer. A favourable response to chemotherapy is indicated by the high value of Ki67. Still, it is not a well established biomarker because of the subjective interpretation involved in its assessment. Programmed Death Ligand 1 (PD-L1) is a significant protein \cite{90} whose presence indicates a possible target for immune checkpoint inhibitors. If high amounts of PD-L1 are detected in the cancer cells, immunotherapy medicines called "immune checkpoint inhibitors" are suggested to be used. It is the most common immune based biomarker used in current clinical settings. PD-L1 protein expression on tumor or immune cells is often measured with the help of IHC assays \cite{17}. However, PD-L1 assessment is quite challenging due to the inter assay variability issues, different scoring algorithms as well as the tissue sampling and preparation difficulties inherent in all IHC based techniques. 

A comparison of different works dedicated for the prediction of various non-omic biomarkers from routine histopathology images can be found in Table \ref{tab3}. The work presented in \cite{18} aims to discriminate between HER2 positive and HER2 negative breast tumors, using a multistage CNN. The first deep learning model, UNet carried out nucleus detection in the stain separated H\&E images. The subsequent CNN performed the tumor vs non-tumor classification followed by the last network for HER2 status identification. The patient level decision was obtained further to the calculation of proportion of HER2+ nuclei among the total cancerous nuclei. A notable work \cite{20} has proposed to use the concept of tissue fingerprints to overcome the limited availability of annotated datasets.  A network is pretrained to learn features that could pair the left/right halves of pathologic images. Images of tissue microarray (TMA) cores were used for training the tissue fingerprint network which enabled to learn the individual tumors. These fingerprint features were used to classify between the tumors with molecular alterations. The molecular information included the assessment of ER, PR and HER2 status.  The pan-cancer study \cite{77}, which aimed to infer multiple molecular features using a single deep learning model, demonstrated that hormone receptor status could be predicted with an Area Under the Receiver Operating Characteristic curve (AUROC) of 0.82 for ER and 0.74 for PR.

\begin{sidewaystable}
\small
\caption{Non-omic biomarkers from H\&E images}\label{tab3}

\begin{tabular*}{\textwidth}{@{\extracolsep\fill}p{1cm}p{3cm}p{2cm}p{2cm}p{2cm}p{3cm}p{5cm}}
\toprule
Reference & \multicolumn{3}{c}{Dataset Details} & Objective & Performance & Method \\ 
\cmidrule(r){2-4}   
& Dataset & Patients & Size & & & \\  
\midrule
\cite{18} & Warwick, TCGA & - & Training: 26 (Warwick), Testing: 26 (Warwick) and 45 (TCGA) & HER2 status & AUC: 0.82 (Warwick), AUC: 0.76 (TCGA) & UNet for nuclei detection, custom CNN for tumor and HER2 classification \\

\cite{20} & TCGA, Australian Breast Cancer Tissue Bank (ABCTB) & - & TCGA: 939, ABCTB: 2531 & ER, PR, HER2 status & ABCTB: AUC- 0.89 (ER), 0.81 (PR), and 0.79 (Her2) & CycleGAN to normalize staining variations, Resnet34 network pretrained on the ImageNet dataset for classification \\

\cite{19} & HER2SC, TCGA & - & Training :52 (HER2SC), 54 (TCGA) & HER2 status & HER2SC: 83.3\% Accuracy, 86.7\% F-score BRCA: 53.8\% Accuracy, 21.5\% F-score & Multiple Instance Learning CNN for HER2 scores, Tile score aggregation by MLP \\

\cite{21} & TCGA, Australian Breast Cancer Tissue Bank (ABCTB) & TCGA: 939, ABCTB: 2535 & TCGA: 1014, ABCTB: 2535 & ER status & AUC: 0.92 on the test set & Otsu Segmentation, Receptor Net with ResNet-50 feature extractor \\

\cite{77} & TCGA & 1007 & 1070 & ER, PR status & AUC: 0.82 (ER), 0.74 (PR)& resnet18, alexnet, inceptionv3, densenet201 and shufflenet \\

\bottomrule 

\end{tabular*}
\end{sidewaystable}

\begin{sidewaystable}
\ContinuedFloat 
\small
\caption{Non-omic biomarkers from H\&E images}\label{tab3}

\begin{tabular*}{\textwidth}{@{\extracolsep\fill}p{1cm}p{3cm}p{2cm}p{2cm}p{2cm}p{3cm}p{5cm}}
\toprule

Reference & \multicolumn{3}{c}{Dataset Details} & Objective & Performance & Method \\ 
\cmidrule(r){2-4}   
& Dataset & Patients & Size & & & \\  
\midrule
\cite{31} & HEROHE dataset, Yale HER2 Cohort & - & HEROHE: 360 (train), 150 (test), Yale: 191 & HER2 status & Balanced Accuracy: 0.741±0.02 (HEROHE), 0.819±0.05(Yale) & Heirarchical prototype clustering, cross attention in transformer for feature aggregation \\

\cite{22} & TCGA, Yale HER2 dataset & - & TCGA: 187, Yale: 188 & HER2 status & AUC: 0.90 (Yale), 0.81 (TCGA) & Tiling and annotation, Data Augmentation, Color Normalisation, Inception V3 Training \\

\cite{23} & TCGA, Tertiary teaching hospital, Medical laboratory & 1513 & 5408 & ER, PR, HER2 status & AUCs: 0.86 (ER), 0.75 (PR), and 0.60 (HER2) & Inception-v3\\

\cite{37} & BCCA and MA31 TMA Dataset & 3376, 275 & 10128 TMA images, 515 TMA images & PD-L1 status & AUC: 0.915 (BCCA), 0.854 (MA31) & Modified ResNet \\

\cite{36} & Clinseq study dataset & 126 & 126 & Ki67 score & Spearman Correlation: 0.527 (weak labels), 0.546 (registered labels), 0.428 (training on transformed images), 0.517 (predicting on transformed images) & Regression CNN \\
\bottomrule 
\end{tabular*}
\end{sidewaystable}

\begin{sidewaystable}
\ContinuedFloat 
\small
\caption{Non-omic biomarkers from H\&E images}\label{tab3}

\begin{tabular*}{\textwidth}{@{\extracolsep\fill}p{1cm}p{3cm}p{2cm}p{2cm}p{2cm}p{3cm}p{5cm}}
\toprule

Reference & \multicolumn{3}{c}{Dataset Details} & Objective & Performance & Method \\ 
\cmidrule(r){2-4}   
& Dataset & Patients & Size & & & \\  
\midrule

\cite{25} & TCGA,  HER2 Contest challenge (HRE2C) and Nottingham University Hospital (Nott-HER2) dataset & 709, 85, 504 & 709, 85, 504 & HER2 status & AUC > 0.75 (TCGA), 0.80 (independent test sets) & Nuclear composition, nuclear morphology and DAB density estimates features with Graph neural network \\

\cite{26} & TCGA,  a proprietary dataset by a private clinical data provider (BioIVT), & - & 648 (ER), 648 (PR) and 560 (HER2) for test set & ER, PR and HER2 status & Accuracy: 87\% (ER), 83\% (PR) and 87\% (HER2) & Resnet34,  Densenet121 \\

\cite{33} & FinProg patient series, FinProg validation series and FinHer clinical trial & 2936, 565, 1010 & 1047 TMA spots, 712 & ERBB2 status & AUC: 0.70 (FinProg), 0.67 (FinHer) & Squeeze and excitation CNN architecture \\

\cite{34} & IHC4BC Dataset & 50 & 240 & ER, PR, HER2, ki67 & AUC above 90 with ViT &CLAM MIL for weak supervision \& Vision Transformer for strong supervision approaches \\

\cite{35} & National Cancer Center, South Korea & 401 & 401 & ER, PR, HER2, ki67, AR & AUC: 0.88 (ER), 0.89 (PR), 0.75 (HER2), 0.86(Ki67), 0.91(AR) & Self supervised learning approach with Simple Framework for Contrastive Learning of Visual Representations (SimCLR) \\

\bottomrule 
\end{tabular*}
\end{sidewaystable}

A weakly supervised approach based on multiple-instance learning (MIL) was proposed in \cite{19} for the classification of HER2 expression. A preprocessing stage was incorporated for the automatic invasive tissue segmentation in order to identify the tumoral regions, and filter the necessary tiles to be fed to the classification stage. The CNN model was pretrained with the corresponding IHC stained images and resulted in 83.3\% classification accuracy on the HER2SC test set and 53.8\% on the BRCA test set. Another multiple instance learning based framework for detecting ER can be found in \cite{21}. The proposed attention based deep neural network, Receptor Net learns from a set of whole slide images with slide level labels. It assumes that a positive slide label indicates there are atleast few regions that has discriminative features for ER-positivity. Conversely, a negative label indicates a complete absence of ER-positive regions. A ResNet-50 feature extractor is used and the aggregate feature vector computed from a random set of tiles selected from the WSI is fed to a decision layer which predicts the ER status for the test WSI. Reference \cite{31} also proposes a weakly supervised approach based on MIL for the detection of HER2 status. A contrastive learning based feature extractor trained on millions of pathologic images is utilized to extract the morphological information from the WSI patches. The significant contribution of the work is the hierarchical prototype clustering module that captures the morphological patterns/ phenotypes across all the slides using a 2 stage clustering process. Cross attention mechanism within the transformer architecture is utilized to combine individual patch features with phenotype embeddings.  Ultimately, this attention mechanism enhances the model's ability to focus on regions of the slide that are most informative for HER2 prediction and hence avoiding manual ROI annotation.

Manually annotated tiles were used for training the deep learning algorithm in \cite{22} to predict HER2 status. Comparative analysis was carried out to demonstrate the significance of pathology annotation for targeted feature learning. There was a strong agreement between pathologist annotated ROIs and the regions responsible for computational predictions of HER2 status. The classifier based on Inception v3 architecture yielded an AUC of 0.90 for slide-level HER2 status prediction in the validation set and 0.81 on independent TCGA data. Paul et.al. proposed three independent deep learning models for determining the status of the three biomarkers; ER, PR and HER2 \cite{23}. Each of them was made up of two stages where the first one is an Inception V3 based CNN trained to categorize the patches as belonging to biomarker positive, biomarker negative, or non-tumor. And the second stage utilizes the features extracted in the first stage to do the binary classification and determine the slide level biomarker status. 

A transfer learning architecture based on modified Xception model accomplishes multistage classification for the prediction of HER2 score (0, 1+, 2+, 3+) \cite{24} in the BC Immunohistochemical (BCI) dataset. The dataset had 4870 pathological image patches in total, including 3896 images in the training set and 977 in the test set generated from 51 whole slide images. An extensive comparison of the proposed deep learning model with existing networks including Inception V3, EfficientNetB7, and DenseNet201 was carried out in the work. Grad-CAM was utilized in the work to apply explainability to the proposed method and hence achieve transparency. They obtained better results than their prior work \cite{27} on the same dataset using Inception V3 architecture. The authors further improved their results in future works \cite{28} and \cite{29}, where they attempted different deep learning architectures such as DenseNet201 and DenseNet201-Xception-SIE respectively. They could achieve an accuracy of 97.12\%, precision of 97.15\%, and recall of 97.68\% with the ensemble model based on DenseNet201 and Xception. In this approach, the features extracted by the model were processed through single instance evaluation (SIE) to find different confidence levels and adjust decision boundary in order to solve the issues with class imbalance. Meanwhile, Wang et. al. also proposed another deep learning architecture HAHNet that combines multi-scale features with attention mechanisms for HER2 score classification \cite{30} on the BCI dataset. However, the results were lower than that with the DenseNet201-Xception-SIE model. Another multi-class classifier was introduced recently in \cite{32} which could yield marginal improvement over the existing results. Three base classifiers (Wide ResNet- 50, DenseNet-201 and GoogleNet) were combined in this work using a class-wise weighted average ensemble algorithm. The comparison of algorithms and performance of different HER2 scoring algorithms can be found in Table \ref{tab4}. 

\begin{table}[!h]
\small
\caption{HER2 scoring from H\&E images}\label{tab4}

\begin{tabular*}{\textwidth}{@{\extracolsep\fill}p{1cm}p{1.25cm}p{1.25cm}p{1cm}p{3cm}p{4.5cm}}
\toprule

Reference & \multicolumn{3}{c}{Dataset Details} & Performance & Method \\ 
\cmidrule(r){2-4}   
& Dataset & Patients & Size & &\\  
\midrule

\cite{24} & BCI benchmark & 319 & 51 & Accuracy (0.87), Precision (0.88), Recall (0.86), and AUC score (0.98) & Modified Xception model with global average pooling, batch normalization, dropout and dense layers, with swish activation function \\

\cite{27} & BCI benchmark & 319 & 51 & Accuracy : 85.1\% & Inception V3 model with additional layers\\

\cite{28} & BCI benchmark & 319 & 51 & Accuracy: 93.45\%, Precision: 94.01\%, and Recall: 93.14\% & DenseNet201 pretrained model modified with additional optimization layers\\

\cite{29} & BCI benchmark & 319 & 51 & Accuracy
: 97.12\%, Precision: 97.15\%, and Recall: 97.68\% & Ensemble of DenseNet201 and Xception, Single Instance Evaluation for addressing class imbalance\\

\cite{30} & BCI benchmark & 319 & 51 & Accuracy
: 93.65\%, Precision: 93.67\%, and Recall: 92.46\% & Multi-scale features with attention mechanisms that combines  Convolutional block attention module (CBAM) and Efficient Channel Attention (ECA)\\

\cite{32} & BCI benchmark & 319 & 51 & Accuracy
: 97.84\%, Precision: 96.62\%, and Recall: 97.87\% & Class-wise weighted average ensemble using 3 algorithms (Wide ResNet- 50, DenseNet-201 and GoogleNet)\\
\bottomrule 
\end{tabular*}
\end{table}

In order to address the issue of limited visual field used for prediction in the patch based methods, a graph neural network model is introduced in \cite{25}. This approach operates on a graph at the entire whole slide image level for an accurate prediction of HER2 status. Both cell-level and contextual information are captured using different feature compositions and a graph neural network is used to generate both regional and WSI level predictions. Reference \cite{26} presents the validation of a device 'PANProfiler' intended to predict the ER, PR and HER2 status. The preprocessing steps comprised of 256 $\times$ 256 tile formation from the WSI, background tile removal and a tumor segmentation algorithm to filter out irrelevant tiles prior to the CNN training. The individual tile scores representing the confidence of being positive were aggregated via mean pooling for the slide level prediction by the system. 

Both weakly and strongly supervised approaches were attempted for predicting ER, PR, HER2 and Ki67 in \cite{34}. A large scale H\&E- IHC paired dataset is contributed by the work and experiments were conducted using clustering-constrained Attention Multiple Instance Learning (CLAM) as well as vision Transformers. Corresponding regions in H\&E and IHC whole slides were manually annotated and registered to obtain patches that can be used for strongly supervised algorithms. The work demonstrated the effectiveness of strong supervision in the task of determining the multiple receptor status from routine histopathology images. Another work intended for the prediction of these multiple biomarkers \cite{35} utilizes a different dataset preparation strategy to overcome the lack of enough training samples. Three dimensional tissue whole slide images which make use of multiple focal planes are captured to build the z-stacked dataset. Further to that, a self supervised learning framework based on Simple Framework for Contrastive Learning of Visual Representations (SimCLR) was deployed to extract features from this z-stacked dataset in a label free manner. However, the tiling process have taken care of confining the region of interest to the cancer region alone as annotated by trained pathologists. The proposed model with ResNet-50 feature extractor was compared with ImageNet pretrained ResNet50 to establish the superiority over conventional supervised learning models. Apart from the commonly identified biomarkers, androgen receptor (AR) status is also predicted in this work.

The work \cite{36} focuses on the prediction of the proliferation marker Ki67 scores from a dataset comprising matched H\&E and Ki67 WSIs taken from 126 breast cancer patients.  In contrast to the binary or multi-class classification tasks mentioned so far, this makes use of regression CNNs to predict the percentage of Ki67-positive
cells. The groundtruth Ki67 scores for each IHC slide were prepared by a trained pathologist using QuPath, which is the percentage of Ki67-positive cells across the cancer
ROI. Four independent models were attempted in the work- Training the H\&E tiles with weak labels, Training with local labels obtained further to registration between the H\&E and Ki67 pairs, Predicting on cycle-GAN generated images, and Training on cycle-GAN generated images. The registration based approach worked better than all others yielding the highest Spearman correlation value of 0.546.

A ResNet based deep learning method for the identification of PD-L1 biomarker in TMA images can be found in \cite{37}. The system delivered an AUC of above 0.90 on a test set from the same cohort which it is trained on and above 0.85 on an independent cohort. The work in \cite{33} accomplishes the prediction of gene amplification status of ERBB2 (HER2) conventionally determined by chromogenic in situ hybridization (CISH) with the aid of a weakly supervised CNN. A squeeze-and-excitation CNN architecture was used to train the tissue microarray core (TMA) images. The binary classifier for ERBB2 amplification tested on a held out set of TMA samples as well as another set of whole slide tissue sections yielded an AUC of 0.70 and 0.67  respectively.
Numerous studies have successfully detected common biomarkers like ER, PR, and HER2 from H\&E images. However, there is a notable gap in research on identifying other biomarkers, such as PD-L1 and Ki67. Additionally, HER2 scoring algorithms have only been evaluated within a single cohort (BCI benchmark), and investigations into ERBB2 amplification detection as a potential alternative to CISH are still sparse.

\section{Computational Methods Replacing Traditional Omic Profiling: Biomarkers and Beyond}\label{sec5}

The introduction of omic technologies in breast cancer management has shown great potential in molecular profiling and the discovery of novel biomarkers \cite{38}. Omics approaches are quite powerful in the sense that it offers the possibility of obtaining a large number of molecular measurements within the tissue. Next Generation Sequencing (NGS)  used for DNA and RNA sequencing is a fundamental tool in genomics and transcriptomics. In genomics, it is used for sequencing DNA to identify genetic variations and mutations, while in transcriptomics, it is used to study RNA expression through RNA-sequencing techniques. Mass Spectrometry is the widely used tool in proteomic and metabolomic profiling. 

\subsection{Genomics}\label{subsec2}
BRCA1 and BRCA2 were the first genes that were discovered in genomics that help repair damaged DNA. Mutations in these genes are linked to a higher risk of breast and ovarian cancers. Since BRCA mutations are hereditary, it is crucial to identify them in the scenario of cancer risk assessment and prevention. TP53 is also a significant tumor suppressor gene which is mutated in more than 50\% of tumors \cite{39}. TP53 mutations are common in Triple Negative Breast Cancer (TNBC), which tends to be more aggressive and has a worse prognosis compared to others.

Reference \cite{40} claims to be the first study to predict the BRCA gene mutation in breast cancer from histopathology slides. A binary classifier is designed with ResNet-18 model which delivered an AUCs between 0.55–0.91 for different image magnifications and models. Another research \cite{41} uses a multi-instance attention model for the prediction of BRCA1/2 gene mutation status in multiple cancer types, including breast cancer. Multi modality was introduced by combining the tissue features with cell features and clinical factors, which was then trained using a Random Forest classifier. The later model (AUC: 0.821) outperformed the uni-modal classifier (AUC: 0.756) on the external test set. A comparison of different methods for the detection of genomic biomarkers from histopathological images can be found in Table \ref{tab5}.

\begin{sidewaystable}
\small
\caption{Genomic biomarkers from H\&E images}\label{tab5}

\begin{tabular*}{\textwidth}{@{\extracolsep\fill}p{1cm}p{3cm}p{2cm}p{2cm}p{2cm}p{4cm}p{4cm}}
\toprule
Reference & \multicolumn{3}{c}{Dataset Details} & Objective & Performance & Method \\ 
\cmidrule(r){2-4}   
& Dataset & Patients & Size & & & \\  
\midrule
\cite{40} & Jiangsu Province Hospital of Chinese Medicine
(JSPHCM) and Jiangsu Cancer Hospital (JSCH), Nanjing & 22, 40 & 110, 112 & BRCA1/2 mutation & AUC on JSCH Dataset: 0.774, 0.804, 0.828, and 0.635 for 40x, 20x, 10x, and 5x magnification slides & ResNet with 18 layers \\

\cite{41} & Chongqing University Cancer Hospital
(CUCH), TCGA & 194, 147 & 248, 149 & BRCA1/2 mutation & Internal test set AUC: 0.859, 0.804; External test set AUC: 0.821, 0.756 & Multi-instance attention model (MIAM) with pretrained ResNet 34, MIAM features, cell features,
and clinical features with Random Forest classifier \\

\cite{42} & TCGA, Uppsala & 943, 207 & 943, 207 & TP53 Mutation & AUC: 0.74, 0.77 for ER+, ER- subgroups & CNN with Multiple instance learning classifier \\
\bottomrule 
\end{tabular*}
\end{sidewaystable}

\begin{sidewaystable}
\small
\ContinuedFloat 
\caption{Genomic biomarkers from H\&E images}\label{tab5}

\begin{tabular*}{\textwidth}{@{\extracolsep\fill}p{1cm}p{3cm}p{2cm}p{2cm}p{2cm}p{4cm}p{4cm}}
\toprule
Reference & \multicolumn{3}{c}{Dataset Details} & Objective & Performance & Method \\ 
\cmidrule(r){2-4}   
& Dataset & Patients & Size & & & \\  
\midrule

\cite{43} & TCGA & 659 & 659 & Point mutation, CNA & AUC for Mutation prediction: 0.852(RB1),  0.776(CDH1),  0.768(NF1), 0.740(NOTCH2),  0.729(TP53),  0.682(MAP3K1); AUC for CNA prediction: 0.794(FGFR1), 0.742(EIF4EBP1), 0.732(KAT6A), 0.715(HEY1), 0.693(ZNF217), 0.686(RAB25) & ResNet-101 with an
attention mechanism \\

\cite{44} & TCGA & - & 9754 in total & whole genome
duplications, chromosomal aberrations, driver gene mutation & AUC: 0.87 (TP53 mutation), 0.7 (ERBB2/HER2 amplification), 0.69 (PTEN Deletion) & Modified Inception-V4 architecture \\ 

\cite{45} & MSK IMPACT Dataset, TCGA & 4912, - & 5967, 1122 & Genomic biomarkers & Virchow2- a foundation
model pre-trained on 3 million slides\\
\bottomrule 
\end{tabular*}
\end{sidewaystable}

\begin{sidewaystable}
\small
\ContinuedFloat 
\caption{Genomic biomarkers from H\&E images}\label{tab5}

\begin{tabular*}{\textwidth}{@{\extracolsep\fill}p{1cm}p{3cm}p{2cm}p{2cm}p{2cm}p{4cm}p{4cm}}
\toprule
Reference & \multicolumn{3}{c}{Dataset Details} & Objective & Performance & Method \\ 
\cmidrule(r){2-4}   
& Dataset & Patients & Size & & & \\  
\midrule

\cite{46} & Stockholm, TCGA, TransNeo, Uppsala & 590, 1076, 165, 207 & 590, 1076, 165, 207 & TP53 mutation status & Best AUC: 0.709 (Stockholm), 0.763(TCGA), 0.697(TransNeo), 0.697 (Cohort2) & 9 MIL algorithms \\

\cite{73} & TCGA & 992 & 1061 & Driver single nucleotide variants (SNV) mutations & 26 out of 128 genes had an AUC >0.7, 0.89 for GNAS & ResNet34 \\

\cite{74} & TCGA TNBC, FUSCCTNBC & 143,425 & 143,425 & PIK3CA mutation, CNA, BRCA2 mutation & AUC: 0.78 (PIK3CA), 0.57 to 0.68 (Top 5 frequent focal amplification and deletion), 0.79 (BRCA2) & ResNet-18 \\

\cite{75} & TCGA, Berlin Cancer Image Base & 565 & 565 & Copy number variation (CNV), Somatic mutation & Balanced Accuracy: 0.6 to 0.7 (CNV), Highest prediction scores for CDH1, TP53 for somatic mutation & Bag of words features and kernel based SVM \\

\cite{77} & TCGA & 1007 & 1070 & single gene mutations and oncogenic driver mutations & AUROC: 0.62 - 0.78 for top 8 mutations & Resnet18, Alexnet, Inceptionv3, Densenet201 and Shufflenet \\
\bottomrule 
\end{tabular*}
\end{sidewaystable}
A CNN feature extractor pretrained on ImageNet with multiple instance learning classifier was used for the prediction of TP53 gene mutation status from H\&E stained slides \cite{42}. The model was trained independently on ER positive and ER negative subgroups of the TCGA- BRCA dataset. The validation of prediction ability and prognostic performance was carried out in an external validation cohort from Sweden which showed an overall AUC of 0.76. Qu et. al used a ResNet-101 (pretrained with ImageNet) and a Multi layer perceptron with attention mechanism to predict the mutation and copy number alteration (CNA) of 6 genes \cite{43}. The point mutation status of six genes (RB1, CDH1, NF1, NOTCH2, TP53, and MAP3K1) could be predicted with an AUC of 0.68–0.85. The copy number alterations of another 6 genes could also be predicted with an AUC ranging from 0.69 to 0.79. The utilization of multiscale features with MIL models \cite{46} demonstrated superior performance over features extracted from single resolution histopathological images. The feature vectors of image patches from 20x and 10x magnification levels were aggregated in the proposed methodology for the prediction of breast cancer grade, TP53 status and survival.  Except for the biopsy based TransNEO dataset, multi resolution models achieved the best performance when experimented with 9 different MIL algorithms.

The prediction of somatic PIK3CA mutation and germline BRCA2 mutation in triple-negative breast cancer cases was conducted in \cite{74}. The study also focused on predicting focal copy number alterations. The deep learning framework involved a two-stage approach: the first stage employed a tissue type classifier trained on pixel-level annotated images, while the second stage used a CNN trained on image tiles to target specific predictions. In the study \cite{75}, statistically significant predictions were made for the copy number variations of 5,274 genes. Additionally, the research involved predicting somatic mutations in 23 genes. The work utilized Bag-of-Words features and kernel-based SVM techniques for accomplishing the molecular feature prediction.

The pan cancer study in \cite{44} utilizes the histopathological features for the prediction of multiple molecular information including whole genome duplications, chromosomal aberrations, driver gene mutations. This includes TP53 mutations and copy number variations like ERBB2 (HER2) amplification, which are crucial for breast cancer diagnosis and treatment. The work also identified the deletion of tumor suppressor genes like PTEN, which is linked to poor prognosis. Another pan cancer approach is presented in \cite{45} which make use of a single deep learning model to predict 1,228 genomic biomarkers for 70 different cancer types. Among the 15 most common cancer types, breast cancer had the third most position in the maximum number of genes that could be predicted with an AUC greater than 0.75. This is contributed by 41 and 17 genes in primary and metastatic samples respectively. The different genomic alterations include amplification, deletion and mutation of various genes including but not limited to CDH1, TP53, ERBB2, BRCA1, PIK3CA. The pan-cancer study in \cite{73} predicts the mutational status of key biomarkers, including TP53, GNAS, BAP1, and MTOR. The study incorporates multi-omic biomarkers and employs independently trained deep learning models. TP53, a mutation frequently associated with poor prognosis, was predicted with a high AUC of 0.785. In contrast to this work, the pan cancer study\cite{77} utilizes single deep learning algorithm trained to predict a number of molecular alterations. Mutations of several genes including TP53, KRAS, BRAF, PIK3CA, MTOR, EGFR, MAP2K4 were considered in this work. Among these, mutations of PIK3CA and MAP2K4 were significantly detectable in breast cancer. While there has been substantial research on genomic biomarker detection from routine pathological images, the detection of BRCA mutations, despite their significance, was minimally explored.

\subsection{Transcriptomics}\label{subsec3}
Transcriptomic techniques assess the expression levels of thousands of genes simultaneously through methods like RNA sequencing or microarray analysis, allowing the identification of active genes in breast cancer cells. Multi-gene signatures deliver biological understanding and aid in risk classification for breast cancer. These gene expression patterns can be utilized for molecular subtyping of breast cancer, which influences prognosis and treatment decisions \cite{47}. One such test is PAM50, a 50-gene signature test that classifies breast cancer into four intrinsic subtypes: Luminal A, Luminal B, HER2-enriched, and Basal-like. Additionally, the OncotypeDX test, which evaluates 21 genes, predicts the likelihood of breast cancer recurrence and the potential benefits of chemotherapy for ER-positive, PR-positive, and HER2-negative breast cancer. The MammaPrint test, featuring a 70-gene expression profile, helps to stratify patients into low or high-risk categories for breast cancer recurrence. Unlike the normal transcriptomic techniques which are accomplished by bulk sequencing or single cell sequencing, spatial transcriptomics \cite{50} allows to do the measurements at spatial resolution. This in turn allows to link the spatial morphology with the related spatial gene expression and aids to understand the tumor heterogenity.

\subsubsection{Image-Driven Gene Expression Profiling}\label{subsubsec1}
The deep-learning algorithm HER2NA, introduced in \cite{48}, is specifically designed to predict gene expression levels from whole slide images. This model was trained using data from approximately 28 different cancer types and demonstrated strong performance in predicting expression levels of genes related to cell-cycle regulation, including CHEK2 and Cyclin E relevant to breast cancer. It effectively predicted approximately 10,000 genes with adjusted p-values below 0.05 specifically for breast cancer. For each predicted gene, the model calculates a score for each tile within the whole slide image (WSI), representing the predicted gene expression for that tile. This scoring method enables the spatial localization of transcriptomic data within the tissue sample. Another pan cancer study can be found in \cite{49}, which work across 9 different cancer types. However, considering the fact that the gene expression profiles vary across cancer types, the model was developed independently for each cancer type. The proposed transformer model SEQUOIA outperformed HER2NA \cite{48} with a drastic increase in the number of significantly well predicted genes from the TCGA test set. A unified model to predict PDL1 expression across multiple cancers uses the Teacher-Student Multiple Instance Learning approach to handle the variability and heterogeneity within and between cancer types \cite{56}. The classification model is intended to discriminate between high and low PD-L1 mRNA expression patterns. The study \cite{76} employs a ResNet-50 architecture that has been pretrained using contrastive self-supervised learning, along with a multi-output regression network, to predict molecular profiles from whole slide images. The system performance is demonstrated using pan-cancer TCGA data and validated with spatial transcriptomic data in breast cancer. A spearman correlation coefficient of 0.722 was obtained for the prediction of 55 tumor microenvironment related genes.

Individual models based on Inception v3 architecture were optimized for each gene across the mRNA transcriptome in breast cancer \cite{54}, which included the prediction of mRNA expression in 17,695 genes. The results indicate that prediction of 9,334 genes was significantly associated with their RNA sequencing values. The predicted variations in intratumoral expression of 76 genes were confirmed through spatial transcriptomics profiling as well.  The study also revealed that genes exhibiting greater expression variance demonstrated slightly enhanced prediction accuracy compared to those with lower variance. This observation is supported by another study \cite{72} that focused on predicting the differential expression of cancer driver genes to assess changes in gene expression between normal and cancerous tissues. Among the 200 cancer driver genes analyzed, the Pearson correlation coefficient for 84 genes was greater than 0.20, while for 39 genes, it exceeded 0.40. Following a similar approach to \cite{54}, a CNN based on Inception V3 is introduced in \cite{55} to detect spatial variations in bulk mRNA and miRNA expression levels, allowing for an analysis of tumor heterogeneity in whole slide images from pathology. 

\begin{table}
\small
\caption{Gene Expression from H\&E images}\label{tab6}

\begin{tabular*}{\textwidth}{@{\extracolsep\fill}p{1cm}p{2cm}p{2cm}p{1cm}p{3cm}p{3cm}}
\toprule

Reference & \multicolumn{3}{c}{Dataset Details} & Performance & Method \\ 
\cmidrule(r){2-4}   
& Dataset & Patients & Size & &\\  
\midrule

\cite{48} & TCGA & 1057 & 1131 &  2902 genes are significantly well predicted & 50-layer ResNet pretrained on the ImageNet \\

\cite{49} & TCGA, CPTAC & 1059, 106 & 1130, 106 & 11,069 well predicted genes, 8587 well predicted genes & Adaptation of vanilla Vision Transformer (ViT) architecture \\

\cite{51} & Spatial transcriptomics dataset,  10x Genomics Spatial Gene Expression dataset & 23 & 68 & Positive correlation between predicted and experimental expression values in 102 of the 250 genes, in more than 20 of 23 patients & DenseNet-121 model with pre-trained ImageNet \\

\cite{52} & Spatial transcriptomics dataset & 23 & 68 & 237 genes identified with positive correlation & Auxiliary network with 4 SOTA deep learning algorithms (ResNet101, Inception-v3, EfficientNet and vision transformer) \\

\cite{53} & Hospital in Japan & 2 & 8 & Pearson correlation coefficient- ESR1: 0.588 (± 0.025), ERBB2: 0.424 (± 0.050),
MKI67: 0.219 (± 0.041) & Adaptation of VGG16 \\
\bottomrule 

\end{tabular*}
\end{table}

\begin{table}
\small
\ContinuedFloat
\caption{Gene Expression from H\&E images}\label{tab6}

\begin{tabular*}{\textwidth}{@{\extracolsep\fill}p{1cm}p{2cm}p{2cm}p{1cm}p{3cm}p{3cm}}
\toprule

Reference & \multicolumn{3}{c}{Dataset Details} & Performance & Method \\ 
\cmidrule(r){2-4}   
& Dataset & Patients & Size & &\\  
\midrule

\cite{55} & TCGA & - & 761 & Average AUC: 0.5(ERBB2), 0.58(ESR1), 0.74(MK167) & Inception V3 \\

\cite{54} & ClinSeq-BC, TCGA, ABiM & 270, 721, 350 & 270, 721, 350 & predicted expression of 9,334 (52.75\%) genes was significantly correlated with measured levels(Spearman correlation, FDR-adjusted P < 0.05) &  Inception V3 regression model\\

\cite{56} & TCGA & - & 266 & AUC for TNBC: 0.54 (FFPE slides), 0.64(Fresh Frozen slides) & ResNet34 with ImageNet pretrained parameters in Teacher-Student collaborated Multiple Instance Learning framework\\

\cite{57} & HER2+ breast cancer dataset & 8 & 36 & Average R: 0.31(GNAS), 0.27(FASN),0.27(MYL12B), 0.26(SCD) for the top 4 predicted among 785 genes & Modified Vision Transformer \\

\cite{73} & TCGA & 992 & 1061 & 18 out of 28 biomarkers with high AUC & ResNet34 \\

\cite{74} & TCGA TNBC, FUSCCTNBC & 143,425 & 143,425 & AUC: 0.78(CD274), 0.71(PDCD1), 0.72(CD8A), 0.70(CD8B) & ResNet-18 \\
\bottomrule 
\end{tabular*}
\end{table}

\begin{table}
\small
\ContinuedFloat
\caption{Gene Expression from H\&E images}\label{tab6}
\begin{tabular*}{\textwidth}{@{\extracolsep\fill}p{1cm}p{2cm}p{2cm}p{1cm}p{3cm}p{3cm}}
\toprule

Reference & \multicolumn{3}{c}{Dataset Details} & Performance & Method \\ 
\cmidrule(r){2-4}   
& Dataset & Patients & Size & &\\  
\midrule
\cite{59} & HER2+ breast cancer dataset & 8 & 36 & Average R: 0.44(FN1), 0.44(SCD), 0.35(IGKC), 0.40(FASN) for the top 4 predicted among 785 genes & Transformer and graph attention networks \\

\cite{60} & Xenium (10x Genomics) Breast Cancer Dataset, Spatial transcriptomics dataset & 23 & 68 & R~0.5 for key genes like FASN, POSTN, IL7R; AUC = 0.94 on ROC curve & Customized Vision Transformer Model\\

\cite{69} & TCGA, Hospital from Australia & - & 495,498 TMA & R = 0.82 across patients,  0.29 across genes for TCGA & EfficientNet, RegNet, DenseNet, Inception, ResNet models \\

\cite{72} & TCGA, CPTAC & - & 153 & Average R of all test genes: 0.185 (P-value<0.01) & Adversarial Contrastive Learning (AdCo) to extract tile-level features and aggregated them with Gated Attention Pooling \\

\cite{75} & TCGA, Berlin Cancer Image Base & 565 & 565 & 7,076 genes are predictable with balanced accuracies between 0.6 and 0.76 & Bag of words features and kernel based SVM \\
\bottomrule 
\end{tabular*}
\end{table}

\begin{table}[!h]
\small
\ContinuedFloat
\caption{Gene Expression from H\&E images}\label{tab6}
\begin{tabular*}{\textwidth}{@{\extracolsep\fill}p{1cm}p{2cm}p{2cm}p{1cm}p{3cm}p{3cm}}
\toprule

Reference & \multicolumn{3}{c}{Dataset Details} & Performance & Method \\ 
\cmidrule(r){2-4}   
& Dataset & Patients & Size & &\\  
\midrule

\cite{76} & TCGA & - & 12592 from all cancer types & R: 0.722 & Contrastive self-supervised learning-based (RetCCL) pretraining and ResNet-50 architecture \\

\cite{58} & HER2+ breast cancer dataset & 8 & 36 & Average R: 0.39(FN1), 0.37(GNAS), 0.35(SCD), 0.33(MYL12B) for the top 4 predicted among 785 genes & Transformer and graph neural networks \\
\bottomrule 
\end{tabular*}
\end{table}

The hist2RNA method presented in \cite{69} was intended to predict the expression of 138 genes. And the method achieved the highest median correlation of 0.29 across genes, and showed statistically significant results for 105 genes. The system was validated on an external tissue micro array (TMA) dataset as well. The binary classifier in \cite{73}, used for identifying the under- or over-expression status of key genes such as ESR1, HER2, PR, and PIK3CA, achieved satisfactory performance, with a high AUC for 18 out of 28 biomarkers. The CD274 mRNA expression (encoding PD-L1) was predicted from WSI images in \cite{74}. Additionally, the expression of other biomarkers, such as PD-1 (based on PDCD1 mRNA) and CD8 (based on CD8A and CD8B mRNA), was also predicted. The work \cite{75} experimented with the prediction of 20,530 genes and found 7,076 to be predictable with balanced accuracies ranging from 0.6 to 0.76.

The work \cite{51} accomplishes the prediction of local gene expression from histopathology images with the help of a deep learning network trained on spatial transcriptomics data. The proposed model ST-Net can predict the spatial variation in the expression of around 102 genes, which includes several breast cancer biomarkers as well. An enhancement over these results is demonstrated in \cite{52}, where an EfficientNet-b0 model with an auxiliary network predicts 237 genes showing positive correlation. Additionally, 24 of these genes achieved a median correlation coefficient exceeding 0.5, compared to the top 5 genes in \cite{51}, which had correlation coefficients around 0.3. The vision transformer-based model in \cite{57} utilized a dataset of 36 tissue sections to demonstrate the prediction of super-resolution gene expression from HER2+ breast cancer images. A comparison with their ST-Net implementation demonstrated that their proposed method, HisToGene, outperformed ST-Net across multiple metrics. The same dataset is utilized in \cite{58} to validate the proposed deep learning model to predict RNA-sequence expression from histology images. The spatial relationships within the image and neighboring patches are captured using a transformer and graph neural network modules and the extracted features are then used for gene expression prediction. A self-distillation mechanism is also incorporated to enhance the model’s learning efficiency. The results were further improved with THItoGene \cite{59}, with a suitable framework to provide comprehensive representation of the complex relationships between images, spatial location, and gene expression. It utilizes a dynamic convolution module with a kernel that adjusts its size based on spatial locations. The vision transformer models in \cite{60} were trained to predict spatial gene expression with single cell resolution. A 40\% improvement in the predictive accuracy over ST-net models was claimed in the work.

Monjo et al. predicts spatial transcriptome profiles from H\&E-stained images using VGG16  networks \cite{53}. The study involved 24 genes including three breast cancer marker genes (MKI67, ESR1, ERBB2) and another 21 breast cancer-related microenvironment marker genes. The predicted gene expression (e.g., ESR1) was correlated with immunohistochemistry results to validate predictions.  Moreover, the work utilized 18,451 genes for the gene set enrichment analysis (GSEA) to identify highly predictable gene sets. This dataset included genes expressed in both tumor and microenvironmental regions.

\subsubsection{Image-Driven Molecular Subtyping}\label{subsubsec2}

\begin{table}
\small
\caption{Molecular subtyping from H\&E images}\label{tab7}

\begin{tabular*}{\textwidth}{@{\extracolsep\fill}p{1cm}p{2cm}p{2cm}p{1cm}p{3cm}p{3cm}}
\toprule

Reference & \multicolumn{3}{c}{Dataset Details} & Performance & Method \\ 
\cmidrule(r){2-4}   
& Dataset & Patients & Size & &\\  
\midrule

\cite{62} & TCGA & 1097 & 1142 & WSI level Accuracy:66.1\%, Patient level Accuracy:67.27\% & Inception V3 and Multiclass SVM \\

\cite{63} & Xiangya Hospital & 1254 & 1254 & Accuracy: 64.3\%, Precision: 65.1\% & ResNet 50 \\

\cite{64} & TCGA, 6 private cohorts & - & 3672 & Maximum AUC for TCGA: 0.803 & ResNet 18 pre-trained on H\&E WSI \\

\cite{65} & 3 Hospitals from china & 603 & 603 & AUC: 0.929 (internal validation), 0.900 (external test) & Resnet18 model pretrained on Imagenet \\

\cite{66} & TCGA, Korea University Guro Hospital & 1009, 480 & 1072, 604 & AUC: 0.749 & ResNet 50 with CLAM \\

\cite{67} & TCGA, Taipei General Hospital dataset & -, 133 & 388, 233 & Maximum slide-wise prediction accuracy: 91.3\% (ResNet101) & VGG16, ResNet50, ResNet101, and Xception models \\

\cite{68} & spatial transcriptomics dataset, TCGA & 23,67 & 68,131 & Accuracy: 80\%(ST data), 73.91\%(TCGA)& VGG16, GoogleNet, Resnet18, Resnet50 and Densenet121 models \\

\cite{70} & TCGA, CPTAC, HER2-Warwick & - & 980, 382, 71 & F1-score: 0.727, precision: 0.741 & Inception V3 for tumor/non tumor classification, ResNet-18 for subtyping \\
\bottomrule 
\end{tabular*}
\end{table}

\begin{table}
\small
\ContinuedFloat
\caption{Molecular subtyping from H\&E images}\label{tab7}
\begin{tabular*}{\textwidth}{@{\extracolsep\fill}p{1cm}p{2cm}p{2cm}p{1cm}p{3cm}p{3cm}}
\toprule

Reference & \multicolumn{3}{c}{Dataset Details} & Performance & Method \\ 
\cmidrule(r){2-4}   
& Dataset & Patients & Size & &\\  
\midrule

\cite{69} & TCGA, Hospital from Australia & - & 495, 498 (TMA) & accuracy: 56\% , F1 score of 55\% & Random Forests,
MultiLayer Perceptron, Linear Discriminant Analysis and Logistic Regression with soft voting \\

\cite{73} & TCGA & 992 & 1061 & AUC: 0.752 & ResNet34 \\

\cite{77} & TCGA & 1007 & 1070 & AUC: 0.78 & Resnet18, Alexnet, Inceptionv3, Densenet201 and Shufflenet \\
\bottomrule 
\end{tabular*}
\end{table}

The PAM 50 molecular subtyping, which is defined by mRNA expression of 50 genes have improved prognostication capabilities compared to other genomic signatures \cite{61}. A method for breast cancer intrinsic molecular subtype classification using deep learning on H\&E-stained WSIs can be found in \cite{62}. The patch-based classifier turned out to be an alternative approach for detecting intratumoral heterogeneity, offering prognostic value. In \cite{63} and \cite{64}, ResNet models are applied to classify the four molecular subtypes of breast cancer. Both studies implement MIL; however, the latter study leverages a specialized backbone pre-trained on H\&E images, yielding better performance than the standard ImageNet backbone. Huang et al. also applied the ResNet 18 model pre-trained on ImageNet, using MIL for feature extraction \cite{65}, focusing on classifying luminal and non-luminal subtypes. Their proposed radiopathomic model featured a Co-Attention module to merge features from pathology and ultrasound images, achieving notable performance with AUC values of 0.929 and 0.900 for the internal and external test sets, respectively. The study in \cite{66} also relies on ImageNet pretrained ResNet 50 with the weakly supervised clustering-constrained attention multiple instance learning (CLAM) approach for molecular subtyping. 

Traditional deep learning models, such as ResNet, VGG, GoogleNet, and DenseNet121, among others, were evaluated for PAM50 molecular subtyping in \cite{67} and \cite{68}. Despite the availability of only slide-level labels, the latter study addressed challenges with noisy patches by employing a patch filtering technique. This approach involved an initial training phase using a spatial transcriptomics dataset to identify relevant patches associated with ESR1, ESR2, PGR, or ERBB2 expression. The patch filtering strategy proved effective in enhancing the model’s performance.

The study in \cite{69}, which focused on gene expression prediction, reported that 32 genes were predicted with a coefficient of determination R² above 0.1, 17 of which belonged to the PAM50 panel. This motivated further experiments on molecular subtyping using the predicted PAM50 gene set, achieving an accuracy of 56\% and an F1 score of 55\% in breast cancer subtype classification. H\&E slides from three distinct datasets were used in \cite{70} to develop and validate a molecular subtyping framework. The algorithm followed a two-stage approach: the first stage employed a binary deep learning classifier to distinguish between tumor and non-tumor regions. For the second stage, four separate binary classifiers were implemented using a One-vs-Rest (OvR) strategy to simplify the four-class classification task. The scores from the four OvR CNNs were utilized as input features for training an XGBoost model, which predicted the molecular subtype of the whole-slide images. PAM50 subtyping using ResNet 34 achieved an AUC of 0.752, with strong performance for the Basal subtype (AUC of 0.871) in \cite{73}. The pan-cancer study \cite{77} employed a single deep learning model to infer various molecular features, achieving an AUROC of 0.78 for PAM50 subtyping.

\subsection{Proteomics}\label{subsec4}

The estimation of protein levels for 223 tumor biomarkers was performed as a multi-task regression task using a weakly supervised contrastive learning approach \cite{71}. The study focused on clinically relevant breast cancer proteins, including ER, PR, HER2, and PD-L1. Protein expression profiles derived from reverse-phase protein arrays (RPPA) were utilized to train the multi-instance learning model. The pan-cancer study \cite{73} assessed the protein-level expression status of key proteins, including ER, PR, HER2, and P53. Among the 32 cancer types analyzed, breast cancer exhibited the highest predictability in proteomic expression, with 37 out of 107 genes achieving an AUC of at least 0.7. 

\begin{table}[!h]
\caption{Proteomic profiles from H\&E images}\label{tab8}

\begin{tabular*}{\textwidth}{@{\extracolsep\fill}p{1cm}p{2cm}p{2cm}p{1cm}p{3cm}p{3cm}}
\toprule

Reference & \multicolumn{3}{c}{Dataset Details} & Performance & Method \\ 
\cmidrule(r){2-4}   
& Dataset & Patients & Size & &\\  
\midrule

\cite{71} & TCGA, CPTAC & 1093, 134 & 1978,  642 & 164 proteins with R >0.2, 16 proteins with R>0.5 & Self-supervised contrastive learning (MoCo v2, SimCLR) \\

\cite{73} & TCGA & 992 & 1061 & 37 out of 107 genes with AUC > 0.7& ResNet34 \\

\cite{75} & TCGA, Berlin Cancer Image Base & 565 & 565 & 19 proteins (statistically significant), max balanced accuracy: 0.65 & Bag of words features and kernel-based SVM \\

\cite{76} & TCGA & - & 12592 in total & R: 0.549 & RetCCL pretraining, ResNet-50 architecture \\

\bottomrule 
\end{tabular*}
\end{table}
The prediction of protein expression is performed as a binary classification task\cite{75}, based on thresholding the protein expression levels across the patient population for the targeted protein.
Among the 190 proteins analysed, 19 proteins had statistically significant predictions with a maximum balanced accuracy of 0.65. The proteomic inference models based on contrastive self-supervised learning \cite{76} evaluated 191 proteins concentrating on tumor-intrinsic and oncogenic processes. The models achieved an average correlation coefficient of 0.549 in their predictions. A comparison of different methods intended for the extraction of proteomic profiles from the routine histopathologial images can be found in Table \ref{tab8}. Proteomic expression prediction, conducted as a binary classification task to indicate underexpression or overexpression as described in \cite{73} and \cite{75}, demonstrated superior performance compared to other regression-based models.

\subsection{Metabolomics}\label{subsec5}
The study in \cite{74} is the first to explore the prediction of metabolomic features from pathological whole slide images. It examines various metabolites and metabolic pathways, with the achieved AUC values ranging from 0.57 to 0.70. The study analyzed TNBC cases from two cohorts, TCGA and FUSCC, achieving a maximum AUC of 0.70 for fatty acid metabolites using ResNet-18 models.

\section{Discussion}\label{sec6}
We have presented several research studies that concentrate on predicting breast cancer biomarkers from routine Hematoxylin and Eosin (H\&E) stained histopathological slides. These studies demonstrate that beyond the identification of individual molecular biomarkers, machine learning techniques hold significant promise in extracting multi-omic biomarkers directly from pathology images. This capability is particularly valuable as it offers the potential to discover novel biomarkers without the need for advanced and often expensive omics-based technologies. The computational models can reveal complex patterns within histological data, facilitating deeper insights into tumor biology and complementing traditional molecular profiling methods. However, the success of these models depends on several critical factors, including the availability and quality of datasets, accurate annotations, model architectures, and explainability techniques. In the following sections, we discuss these aspects in detail. 

\subsection*{\textbf{Dataset availability and Limitations:}}
The availability of large-scale datasets with both histopathological and molecular annotations has been essential for the development of robust predictive models. Many studies have utilized public datasets, such as The Cancer Genome Atlas (TCGA), to build their models, while others rely on proprietary clinical datasets. However, due to ethical concerns surrounding patient privacy, most clinical datasets remain inaccessible, making it challenging to compare models developed using these private datasets.

Given that the assessment of ER, PR, and HER2 status is now part of routine molecular profiling in breast cancer, acquiring such data is relatively straightforward. For instance, the HEROHE dataset\footnote{https://ecdp2020.grand-challenge.org/Dataset/}  was introduced as part of a challenge to predict HER2 status from H\&E-stained whole slide images (WSIs) \cite{78}, providing slide-level annotations indicating positive or negative HER2 status. Another valuable resource is the Yale HER2 cohort\footnote{https://www.cancerimagingarchive.net/collection/her2-tumor-rois/}, consisting of WSIs from Yale School of Medicine, which includes region-of-interest (ROI) tumor annotations. These datasets contribute to the pool of public resources supporting HER2 status prediction through image analysis-based approaches. In addition, the BCI benchmark dataset\footnote{https://bupt-ai-cz.github.io/BCI/}, containing paired H\&E and IHC WSIs \cite{79}, has been widely used in studies focused on HER2 scoring applications. Notable works have exploited this dataset to enhance model performance and reliability in HER2 scoring tasks\cite{24},\cite{27},\cite{28}. Furthermore, two relatively large public datasets specifically designed for the validation of image analysis algorithms targeting ER, PR, HER2, and Ki67 \footnote{https://snd.se/sv/catalogue/dataset/2022-190-1/1}, \footnote{https://ihc4bc.github.io/} assessments has been presented in \cite{30} and \cite{80}.

However, predicting omic biomarkers from H\&E slides presents additional challenges, as it requires datasets that have corresponding omic profiles—such as mutations, RNA expression, metabolomic, or proteomic data. Unlike hormone receptor profiling, these omic technologies are not routinely employed in clinical laboratory settings, nor are they part of standard diagnostic workflows in many cases. This makes data acquisition for omic biomarker prediction more difficult. As a result, most algorithmic validations for omic biomarkers have been conducted using the TCGA database, which provides both histopathological images and multi-omic data. Other datasets with similar multi-omic profiles tend to be relatively small in size \cite{53}, limiting the scope of large-scale model validation.

Different populations (e.g., Asian, Middle Eastern, African, or European) \cite{81} exhibit distinct genetic and molecular profiles that influence the expression of biomarkers like ER, PR, HER2, and other omics-based features. This variability can impact model generalizability and the models trained on predominantly Western datasets may not perform well on datasets from underrepresented populations such as African or middle east cohorts. Many public datasets, such as TCGA and the Yale HER2 cohort, are dominated by data from Western populations, with relatively fewer samples from Asian, Middle Eastern, or African populations. The lack of representation of certain ethnic groups introduces a risk of bias, which may limit the clinical applicability of these models in diverse patient populations. Population specific models or personalized prediction frameworks should be explored to address regional and ethnic differences. For eg:- Developing models specifically tailored for Middle Eastern populations may improve the prediction accuracy of hormonal biomarkers and molecular subtypes relevant to this cohort. Additionally, these models can enhance biomarker discovery, revealing previously unknown genetic or molecular features associated with certain ethnic groups.

\subsection*{\textbf{Annotation Benefits and Challenges:}}
Whole slide images (WSIs) are extremely high-resolution images, often as large as 100,000 × 100,000 pixels, making it necessary to divide them into smaller tiles or patches for analysis. However, many of these patches contain non-informative areas, such as empty regions or non-tumorous tissue, which complicates training models. The process of pixel-wise or region-level annotations by expert pathologists is time-consuming, expensive, and impractical for large datasets, which is why many datasets only provide slide-level labels. In some studies, pathologists have annotated tumor-specific regions using specialized software tools such as QuPath \cite{41}, Aperio ImagescopeR \cite{18}, \cite{22}, and the Automated Slide Analysis Platform \cite{40}. This targeted annotation helps generate information-rich patches, which improve model training by focusing on relevant areas. Automated segmentation algorithms deployed in \cite{18}, \cite{19}, \cite{26}, \cite{54} could distinguish tumor regions from non-tumorous areas, ensuring that irrelevant regions are excluded from the analysis.

Utilizing patch-level annotations, such as in \cite{23}, enables validating models for localized biomarker predictions within specific tumor regions. This localized analysis is particularly important for capturing biomarker heterogeneity, where biomarker expression can vary across different areas of the tumor. For example, patch-level predictions help identify regions with differing HER2 or ER statuses within the same tumor, which can affect treatment decisions. Unlike whole-slide analysis, tissue microarray (TMA) images as used in \cite{82} offer a more focused approach by containing carefully selected tumor regions. This selection minimizes noise from non-tumorous areas, improving the precision of molecular subtyping. The study achieved 77\% accuracy in distinguishing Basal-like from Non-Basal-like tumors and 84\% accuracy in predicting ER status, demonstrating the value of working with representative tumor regions for specific biomarker assessments.

The lack of region-level annotations remains a significant challenge in digital pathology. This challenge can be addressed at the algorithmic level through weakly supervised learning approaches, such as attention-based multi-instance learning (MIL), which has been employed in many studies referenced in this article. MIL treats each WSI as a bag of smaller instances (tiles), with the entire slide assigned an overall label (e.g., ER-positive or negative). The attention mechanism helps the model focus on the most relevant tiles that contribute to the overall label. During training, the attention weights dynamically assign importance to certain tiles, enabling the model to prioritize regions more likely associated with a biomarker (e.g., tiles showing ER-positive patterns). This approach enables the model to learn region-specific patterns without requiring explicit patch-level labels, bridging the gap between weakly supervised learning and region-specific predictions.

While MIL has proven effective in some applications, its performance can vary. For example, a study \cite{34} reported significant performance differences when comparing models trained with patch-level annotations against those using only slide-level labels. Specifically, weakly supervised methods showed limitations in accurately predicting the ER, PR, Ki67, and HER2 statuses. The use of tile-level annotations in \cite{22} significantly improved the HER2 status determination accuracy. This suggests that, despite the advantages of MIL, region-level annotations may still be necessary to achieve optimal performance for certain biomarker predictions. Some studies for molecular subtyping have addressed the issue of noisy patches by employing patch filtering techniques as well \cite{63}, \cite{68}.

\subsection*{\textbf{Deep Learning Models and Architectures:}}
As discussed in the previous section, most studies have adopted a weakly supervised Multi-Instance Learning (MIL) approach to address the challenges associated with the lack of patch-level annotations. Among the various deep learning models, traditional architectures such as ResNet have been widely used across numerous works. In addition, several studies have conducted comparative analyses by utilizing different deep learning architectures, including DenseNet and Inceptionv3. However, the Graph Neural Network (GNN) model proposed in \cite{25} stands out from these conventional CNN-based architectures. Unlike ResNet, DenseNet, or AlexNet, the GNN model takes advantage of graph structures to capture complex relationships and contextual dependencies within Whole Slide Images (WSIs). This capability enables the GNN to generate more accurate and nuanced predictions of HER2 status, effectively accounting for the intricate spatial patterns found in biological tissues—something that traditional CNNs may struggle to achieve.

One critical component in the development of deep learning models is pretraining. In most cases, ImageNet pretraining has been the standard choice for initializing models before fine-tuning them for specific tasks in pathology \cite{21}, \cite{24}, \cite{32}, \cite{42}. However, recent works have begun to explore the advantages of pretraining models on domain-specific data. For example, the study in \cite{49} demonstrated that transformer models pretrained on histopathology data from normal tissues (excluding breast tissues) outperformed ImageNet-pretrained models in gene expression prediction. This highlights the benefit of using data from the same modality for pretraining, as it enables the model to capture tissue-specific patterns more effectively.

A notable example of this trend is found in \cite{72}, which employed self-supervised pretraining on large-scale unannotated histopathology tiles extracted from WSIs for differential gene expression prediction. This approach allows the model to learn morphological patterns directly from histopathology images, providing more relevant feature representations for downstream tasks compared to features learned from general-purpose datasets like ImageNet. By focusing on domain-specific histopathology tiles, the model becomes better aligned with the biological context, leading to improved performance in tasks like gene expression prediction. Another study \cite{64} also demonstrated that pretraining on H\&E-stained images achieved better performance than ImageNet pretraining for molecular subtype prediction. These findings suggest that domain-specific pretraining using histopathology images can significantly enhance model performance for pathology-related tasks by enabling the network to learn features tailored to the complexities of tissue morphology.

\subsection*{\textbf{Explainability and Interpretability of Models:}}
In contrast to cancer diagnosis tasks, it is not feasible to manually inspect standard histopathology slides under a microscope for the relevant molecular profiling. Experienced pathologists must rely on IHC-stained images to determine ER, PR, and HER2 status, as the relevant features for these assessments are not visible in H\&E images. Computer vision methods can extract information that is beyond human perception. Consequently, the decisions made by the system often remain opaque unless clarified through model explainability \cite{85}. This visualization is essential for understanding the morphological patterns associated with biomarker predictions, aiding pathologists in validating the model's conclusions.

Weakly supervised learning models, including those based on multiple instance learning (MIL), typically produce slide-level predictions from instance-level data. Instead of averaging the predictions of all patches in the classifier head, the design can maintain the class probabilities of each tile \cite{31}, allowing the model to highlight specific regions. These spatially mapped scores are visualized as heatmaps, with more confident regions emphasized, enabling clinicians to see which areas most influenced the predictions. A heatmap is generated in \cite{22} using tile-predicted probabilities to visualize the model's predictions. The tile-level heatmaps showed strong agreement with tile-level pathologist annotations while employing a three-class classifier. The model generated heatmaps are used to highlight the regions of the slide that contributed most to the HER2 predictions in \cite{25} and \cite{31}. High positive or negative ER predictions are accompanied by heatmaps in \cite{20} for regional visualizations. Additionally, visualizing the t-SNE embeddings of whole slide image fingerprints provides valuable insights into how a deep learning model organizes and interprets tissue patterns. It helps in identifying clusters of similar samples and validating learned representations, making it a useful tool in medical AI workflows \cite{21}, \cite{37}. Training and Validation Attention Consistency (TAVAC) in attention-based deep learning models measures the consistency of attention weights between the training and validation phases. The reliability of the model in focusing meaningful areas for prediction is ensured if it exhibits similar attention to relevant regions during both training and validation, thereby enhancing interpretability and trustworthiness. The TAVAC score uses Pearson correlation to quantify the consistency of the attention maps in \cite{60} for spatial gene expression prediction.

An explainable AI approach utilizing Layer-Wise Relevance Propagation (LRP) provides spatial insights into the relationships between molecular and morphological features \cite{75}. This helps to identify correlations between spatial features such as tumor cells and molecular markers like gene expression. The model’s interpretability in \cite{23} was enhanced through multiple methods: TCAV (Testing with Concept Activation Vectors) to test if model features aligned with histologic concepts, patch clustering to identify shared morphological patterns linked to predicted biomarkers, and SmoothGrad to generate pixel-based saliency maps that highlight the most influential regions for biomarker predictions. Gradient-weighted Class Activation Mapping (Grad-CAM) is employed in \cite{24} and \cite{28} to create visual explanations for the proposed deep learning models. It operates by computing the gradients of the target class score with respect to the feature maps of a convolutional layer, emphasizing the regions in the input image that have the most influence on the prediction. The Shapley Additive exPlanations (SHAP) algorithm is utilized in \cite{41} to enhance the explainability of the models.

\subsection*{\textbf{Future Directions and Clinical Impact:}}
Molecular profiling from histopathology images holds significant potential in clinical applications. Intra-tumoral heterogeneity (ITH) arises when different regions within the same tumor exhibit varying levels of protein expression or gene amplification. This biomarker heterogeneity is often linked to treatment resistance and disease progression. For instance, patients with heterogeneous HER2 expression may respond poorly to targeted therapies, and it has been reported that HER2 ITH occurs in 40\% of breast cancers \cite{83}. Detecting such heterogeneity is essential for guiding personalized treatment plans, such as determining eligibility for anti-HER2 therapy. While techniques like single-cell profiling and spatial transcriptomics can identify these variations, they are expensive and limited to covering only a few thousand cells. In contrast, digital profiling of molecular features offers a cost-effective solution. Notably, super-resolution techniques in spatial transcriptomic profiling \cite{53} enhance the visualization of tumor invasion regions that are not detectable with raw data alone. Additionally, superior imputation of tissue sections to predict gene expression can further reduce costs and enhance data completeness.

Many studies focused on biomarker prediction, gene expression profiling, and proteomic analysis have adopted a pan-cancer approach, where a single model is used to extract features common across multiple cancer types. Some studies, however, employ separate deep learning models for specific tasks \cite{73}. Pan-cancer models assume that different cancers share similar morphologies, but this assumption overlooks the broad morphological diversity observed across tumors. For example, an analysis of 11 cancer types revealed that not all tumors exhibit distinct patterns of PD-L1 expression, making it challenging to compare them directly to PD-L1-relevant tumors \cite{56}. When cancers display unique morphological characteristics and varying levels of PD-L1 expression, a single model may struggle to generalize effectively, leading to poor performance for tumors with features that diverge from the model’s assumptions. In such cases, developing specialized models tailored to specific genes could yield more accurate predictions by directly linking gene expression to tumor morphology.

The enrichment of E-cadherin (CDH1) mutations in lobular breast cancer and its strong predictive score reflects the deep link between morphology and specific molecular alterations \cite{75}. The high predictive performance stems from distinct structural patterns caused by CDH1 mutations, which are more easily detected by histology-based machine learning models. This suggests that genes with higher fold changes, those driving significant molecular and phenotypic transformations are more amenable to prediction using computational models. This insight highlights the potential for future research to focus on gene expression prediction from histopathology images, particularly for cancer-related genes. The detection of ERBB2 amplification from H\&E slides has also received limited attention. Employing multimodal approaches, as in \cite{72}, could improve biomarker identification performance. Furthermore, metabolomic and proteomic profiling derived from histology images is an underexplored area in breast cancer research, offering a promising avenue for future investigations.

\section{Conclusion}\label{sec7}
The integration of AI in digital pathology offers transformative potential for improving breast cancer prognosis and tailoring treatment plans through biomarker identification. In our study, we conducted an in-depth review of recent research examining AI’s role in extracting biomarkers from histopathology images. While protein biomarkers have been well-explored, we found that research on omic biomarkers—such as genomic, transcriptomic, proteomic, and metabolomic markers is still relatively limited. This gap largely stems from limited dataset availability for developing and validating algorithms. Addressing key challenges, including model interpretability, dataset diversity, and annotation feasibility, will be essential to bridge the gap between research and clinical implementation. Our findings emphasize digital pathology's growing role in cancer care by enabling biomarker identification crucial for personalized treatment. Future advancements in AI-based approaches and the development of standardized protocols for biomarker validation could position AI-enabled pathology as a foundational element in precision oncology.

\section{Acknowledgements}

This publication was made possible using a grant from the Qatar National Research Fund through Academic Research Grant (ARG) ARG01-0513-230141.







\end{document}